\documentclass[pra,twocolumn,showpacs,superscriptaddress,notitlepage,nofootinbib,amssymb,preprintnumbers]{revtex4-1}

\usepackage{amsmath,amssymb,amsthm}
\usepackage{braket}
\usepackage{empheq}

\usepackage[pdfstartview=FitH]{hyperref}
\hypersetup{
    colorlinks=true,       
    linkcolor=cyan,          
    citecolor=magenta,        
    filecolor=magenta,      
    urlcolor=cyan,           
    runcolor=cyan
}

\newcommand{\bes} {\begin{subequations}}
\newcommand{\ees} {\end{subequations}}
\newcommand{\beq}{\begin{equation}}
\newcommand{\eeq}{\end{equation}}
\newcommand{\ba}{\begin{eqnarray}}
\newcommand{\ea}{\end{eqnarray}}

\newcommand{\mrp}{\mathrm{p}}

\newcommand\Tr{\mathrm{Tr}}
\newcommand{\ketbra}[1]{|{#1}\rangle\langle#1|}

\newtheorem*{theorem*}{Theorem}

\def\a{\alpha}
\def\b{\beta}
\def\g{\gamma}
\def\d{\delta}
\def\e{\epsilon}

\def\o{\omega}

\def\G{\Gamma}

\begin{document}
	\title{Error Suppression for Hamiltonian Quantum Computing in Markovian Environments}
	
	\author{Milad Marvian}
	\affiliation{Department of Electrical Engineering, University of Southern California, Los Angeles, California 90089, USA}
	\affiliation{Center for Quantum Information Science \&
		Technology, University of Southern California, Los Angeles, California 90089, USA}
	
	\author{Daniel A. Lidar}
	\affiliation{Department of Electrical Engineering, University of Southern California, Los Angeles, California 90089, USA}
	\affiliation{Center for Quantum Information Science \&
		Technology, University of Southern California, Los Angeles, California 90089, USA}
	\affiliation{Department of Physics and Astronomy, University of Southern California, Los Angeles, California 90089, USA}
	\affiliation{Department of Chemistry, University of Southern California, Los Angeles, California 90089, USA}

\begin{abstract}
Hamiltonian quantum computing, such as the adiabatic and holonomic models, can be protected against decoherence using an encoding into stabilizer subspace codes for error detection and the addition of energy penalty terms. This method has been widely studied since it was first introduced by Jordan, Farhi, and Shor (JFS) in the context of adiabatic quantum computing. Here we extend the original result to general Markovian environments, not necessarily in Lindblad form. We show that the main conclusion of the original JFS study holds under these general circumstances: assuming a physically reasonable bath model, it is possible to suppress the initial decay out of the encoded ground state with an energy penalty strength that grows only logarithmically in the system size, at a fixed temperature. 
\end{abstract}
\maketitle
	
\section{Introduction}	

Hamiltonian quantum computing includes the adiabatic and holonomic models. Adiabatic quantum computing (AQC) \cite{farhi_quantum_2000} is a model that can achieve universality \cite{aharonov_adiabatic_2007,OliveiraGadget,PhysRevLett.99.070502,Breuckmann:2014,Gosset:2014rp,Lloyd:2016} and appears promising for near future large scale realization (for a review see Ref.~\cite{Albash-Lidar:RMP}). In AQC, the computation is performed using a time-dependent Hamiltonian that evolves slowly from an initial Hamiltonian with a known and easily preparable ground state, to a final Hamiltonian whose ground state is unknown and encodes the desired result. The adiabatic theorem guarantees that the final state will be close to the ground state of the final Hamiltonian if the evolution is sufficiently slow \cite{Jansen:07}. Holonomic quantum computing (HQC) is another universal model, wherein quantum gates are performed as holonomies (non-Abelian geometric phases) in the degenerate ground eigensubspace of the system Hamiltonian \cite{HQC,Duan:2001ff,Recati:02}.

Unfortunately, AQC lacks a theory of fault tolerance, unlike all other universal models of quantum computation \cite{Lidar-Brun:book}. The first scheme to suppress the detrimental effect of the bath on AQC \cite{childs_robustness_2001,PhysRevLett.95.250503,Aberg:2005rt,ashhab:052330,PhysRevA.75.062313,amin_decoherence_2009,PhysRevA.80.022303,Sarovar:2013kx,Albash:2015nx} was proposed by Jordan, Farhi, and Shor (JFS) \cite{jordan2006error}. In this scheme, a stabilizer subspace code that can detect the errors introduced by the system-bath interaction Hamiltonian is chosen, and the system Hamiltonian is encoded using the logical operators of the same code. Adding a penalty Hamiltonian breaks the induced degeneracy and stabilizes the  computation in the code-subspace, 
 while any excitation out of this subspace is penalized. The short time performance of this scheme was investigated for a specific Markovian model in Ref.~\cite{jordan2006error} and also for a general non-Markovian bath in \cite{Bookatz:2014uq}, where numerical simulations were used to extend the study beyond the short time limit. For a general but local non-Markovian bath in the regime of weak coupling to the bath, it was shown that, modulo a unitary rotation in the codespace due to the Lamb shift, the same scheme can result in an exponential suppression of decoherence \cite{Marvian:2016aa}. Generalizations to subsystem codes have been proposed \cite{Jiang:2015kx}, and theoretically proven to work \cite{Marvian-Lidar:16}.
Variants of the JFS scheme tailored to current experimental quantum annealing \cite{DWave}, where encoding of the initial Hamiltonian is not possible, have also been proposed and studied \cite{PAL:13,MNAL:15,vinci2015nested}. 

The effects of decoherence and its mitigation in HQC have also been the subject of intensive study \cite{Solinas:04,Wu:2005aa,Florio:06,sarandy_abelian_2006,sol07,oreshkov_adiabatic_2010,Xu:2012aa}.
While unlike AQC, a theory of fault-tolerance has been developed for HQC \cite{Oreshkov:2009bl,Oreshkov:2009lq}, it is of interest to develop less demanding alternatives, such as the error suppression strategy we consider here.

In Section~\ref{sec:2} we show how the results of JFS \cite{jordan2006error} can be extended beyond the specific (photonic bath) model considered there to arbitrary Markovian dynamics, and beyond protecting pure states to the protection of mixed states in degenerate ground subspaces.
Starting from a master equation derived in Ref.~\cite{ABLZ:12-SI} for a system evolving adiabatically while weakly coupled to bath, we show that the main conclusion of Ref.~\cite{jordan2006error} holds very generally for physically reasonable (i.e., local and thermal) models of the bath and for arbitrary ground state degeneracy: the energy penalty is only required to grow logarithmically in the system size, at fixed temperature. In Section~\ref{sec:3} we show that this result stands even if the Markovian master equation is not in Lindblad form, i.e., is derived without applying the rotating wave (or secular) approximation.

The reason we are interested in Markovian models, despite the fact that general results of a similar nature have already been established for non-Markovian models \cite{Bookatz:2014uq,Marvian:2016aa}, is that Markovian models are special: not only are they widely used \cite{Breuer:2002}, decay in these models (e.g., of the purity) is always exponential \cite{Alicki:87}. This means that they preclude any use of ultra-short time recurrence effects that soften decoherence. In particular, error suppression techniques such as dynamical decoupling \cite{PhysRevLett.100.160506,PhysRevA.86.042333,Ganti:13} or the Zeno effect \cite{PhysRevLett.108.080501} (shown to be formally equivalent to the JFS scheme \cite{Young:13}) are ineffective for Markovian models. In this sense, error suppression for Hamiltonian computation in the presence of a Markovian environment is more challenging than in the non-Markovian case.

\section{Error suppression for general master equations in Lindblad form}
\label{sec:2}

Assuming a time-dependent system Hamiltonian $H(t)$, a general bath Hamiltonian $H_B$, and an interaction Hamiltonian $H_{SB}=\sum_{\alpha}{A_{\alpha} \otimes B_{\alpha}}$, an adiabatic Markovian master equation in Lindblad form \cite{Davies:74,Lindblad:76} can be derived \cite{ABLZ:12-SI}:
\begin{eqnarray}
	\dot{\rho}=-i [H(t)+H_{LS}(t),\rho]+D(t)[\rho]\ ,
	\label{eq:ME}
\end{eqnarray}
where $H_{LS}(t)$ is the Lamb shift, and $D(t)$ denotes the dissipative, i.e., non-unitary part (see Ref.~\cite{Albash:2015nx} for a concise summary and definitions), and we set $\hbar\equiv 1$ throughout. Henceforth we mostly suppress the time-dependence of the various terms for notational simplicity, but it important to remember that all our quantities are explicitly time-dependent. 

Consider the spectral decomposition 
\beq
H = \sum_{l\geq 0} \e_l \Pi_l\ ,
\eeq
i.e., $\Pi_l$ denotes the projection onto the (possibly degenerate) $H$-eigensubspace with energy $\epsilon_l$. The eigenprojectors are orthogonal: $\Pi_l \Pi_{l'} = \delta_{ll'}\Pi_l$. Defining $A_{\alpha}(\omega)=\sum_{\epsilon_{l'}-\epsilon_l=\omega}{\Pi_lA_{\alpha}\Pi_{l'}}$, the dissipator becomes:
\begin{eqnarray}
	D[\rho]=\sum_{\omega}\sum_{\alpha\beta}\gamma_{\alpha\beta}(\omega)[A_{\beta}(\omega)\rho A_{\alpha}^{\dagger}(\omega) \quad \quad\notag \\
	-\frac{1}{2}\{A_{\alpha}^{\dagger}(\omega)A_{\beta}(\omega),\rho\}] \ ,
	\label{eq:diss}
\end{eqnarray}
where the matrix of decay rates 
\beq
\gamma_{\alpha\beta}(\omega)= \int_{-\infty}^{\infty}dt\ e^{i\omega t} \langle \mathcal{B}_{\a\b}(t)\rangle = \g^*_{\b\a}(\o)\ ,
\eeq 
is the Fourier transform of the bath correlation function
\beq
\langle \mathcal{B}_{\a\b}(t)\rangle = \Tr[\rho_B e^{-i H_B t} B_\alpha e^{i H_B t} B_\beta]\ ,
\eeq
where $\rho_B$ is the initial state of the bath.

From now on we assume that the system-bath coupling exhibits a local structure, in the sense that the system operators $A_{\alpha}$ in $H_{SB}=\sum_{\alpha}{A_{\alpha} \otimes B_{\alpha}}$ are $k$-local, with $k$ a constant that is independent of the number of qubits $n$. This guarantees that the interaction Hamiltonian can be expressed in terms of a number of terms that is polynomial in $n$.

\subsection{General expression for the excitation rate after encoding and error suppression}

Assume that the system is initially prepared in the (possibly degenerate) ground subspace of the Hamiltonian $H$, with energy $\e_{0}$, i.e., $\rho(0)=\Pi_0 \rho(0) \Pi_0$.
We are interested in the initial excitation rate out of the ground subspace: 
\beq
R\equiv\partial_t\Tr[\Pi_0{\rho}] |_{t \simeq 0 } = \Tr[\Pi_0 \dot{\rho}(0)] \ ,
\label{eq:R}
\eeq 
where the second equality is proved in Appendix~\ref{app:A}. 
In AQC one is usually interested in the case that the ground state of the initial Hamiltonian is non-degenerate and the initial state is pure. In this case the excitation rate $R$ is proportional to the initial purity decay, with purity defined as $\Tr{\rho^2}$. In HQC the initial state belongs to a degenerate subspace. In Eq.~\eqref{eq:R} we do not assume that the initial state is pure, and later consider the special case when it is [see below Eq.~\eqref{eq:purity-loss}].
 
It is not hard to show (see Appendix~\ref{app:B}) that the dissipative part yields:
\begin{align} 
\label{eq:rate}
&\Tr\{\Pi_0 D[\rho(0)]\} \notag \\
&\quad =-\sum_{\alpha\beta}\sum_{l\neq 0}\gamma_{\alpha\beta}(\e_0-\e_l) \Tr[\rho(0) A_{\alpha}^{\dagger} \Pi_l A_{\beta}  ] \ .
\end{align}
	
We now choose a code $C$ that can detect all the errors (system operators) $A_\a$ in the system-bath Hamiltonian \cite{Knill:1997kx}:
\beq
\forall \a: \, P_C A_\alpha P_C=0\ ,
\label{eq:err-det}
\eeq
where $P_C$ projects onto the code space. We encode the system Hamiltonian using the logical operators of this code, and add a penalty Hamiltonian that has the codespace as its ground-subspace. Such a Hamiltonian can be constructed by summing the stabilizer generators of the code \cite{jordan2006error}. Thus, $H$ is the sum of an encoded computational Hamiltonian $\overline{H}_S$ and a penalty Hamiltonian $H_{\mrp}$:
\ba
H(t)=\overline{H}_S(t) + \eta_{\mrp} H_{\mrp} \ ,
\label{eq:H(t)}
\ea 
where the dimensionless quantity $\eta_{\mrp}>0$ quantifies the strength of the energy penalty, and by construction $[\overline{H}_S(t), H_{\mrp}]=0$. This allows us to choose the $\Pi_l$'s as the simultaneous eigenprojectors of $\overline{H}_S$ and $H_{\mrp}$, and write the eigenvalues of $H(t)$ as
\beq
\epsilon_l(t) = \overline{\o}_l(t) + \eta_{\mrp} \xi_l\ , 
\label{eq:e_l}
\eeq
where $\overline{\o}_l(t)$ and $\xi_l$ are, respectively, the eigenvalues of $\overline{H}_S(t)$ and $H_{\mrp}$.
Let us assume that $H_{\mrp}$ is chosen so that its ground subspace is the codespace, defined by the projection operator 
\beq
P_C=\sum_{l \in C}{\Pi_l}\ ,
\label{eq:P_C}
\eeq 
and that the initial state belongs to the (now definitely degenerate) ground subspace of $H(t)$, i.e., again $\rho(0)=\Pi_0\rho(0)\Pi_0$. 
Since $\Tr(X[Y,X])=0$ for any pair of operators $X,Y$, the unitary part $-i [H+H_{LS},\rho(t)]$ of the master equation~\eqref{eq:ME} does not contribute to the initial excitation rate.\footnote{The effect of the Lamb shift on the codespace is captured by other measures such as the fidelity (see, e.g., Ref.~\cite{Marvian:2016aa}).} 
Moreover, because of the error detection properties of the code [Eq.~\eqref{eq:err-det}] we have 
\begin{align}
\forall l \in C: \Pi_0 A_\alpha \Pi_l= 0\ .
\end{align}
Using the master equation~\eqref{eq:ME} and Eq.~\eqref{eq:rate} we thus have:
\begin{align}
R&=-\sum_{\alpha\beta}\sum_{l \in C^{\perp}}\gamma_{\alpha\beta}(\e_0-\e_l)\Tr[\rho(0)A_{\alpha}^{\dagger}\Pi_l A_{\beta}] \ .
\label{eq:exci-rate}
\end{align}

Note that the matrix $\g$ is positive semi-definite and can be diagonalized by a unitary $U$: $\sum_{\a'\b'} (U^\dagger)_{\a\a'} \g_{\a'\b'} U_{\b'\b}= \d _{\a\b} \g_{\a}$ with positive $\g_{\a}$ (the eigenvalues of $\g$). Introducing new Lindblad operators $F_{\a'}(\o)$ via $A_\a (\o)=\sum_{\a'} U_{\a\a'} F_{\a'}(\o)$ into Eq.~\eqref{eq:exci-rate} we have the following general expression for the excitation rate:

\begin{align}
R&=-\sum_{\alpha}\sum_{l \in C^{\perp}}\gamma_{\alpha}(\e_0-\e_l) \Tr[\rho(0)F_{\a}^{\dagger}\Pi_l F_{\a}] \ .
\label{eq:purity-loss}
\end{align}

Alternatively, when the initial state is a pure state $\ket{\psi_0}$ we can define the excitation rate as $R'\equiv\Tr[\ketbra{\psi_0}{\dot{\rho}(0)}] |$, but it is easy to check that as a result of the encoding we have $R=R'$. This means that the encoding also suppresses the errors induced by the system-bath interaction in the ground subspace, which are logical errors for HQC.

\subsection{The excitation rate scales only polynomially in the system size}
Despite the fact that the sum over $l \in C^{\perp}$ involves exponentially many terms, \emph{the excitation rate scales only polynomially in the system size}. To see this, we first define
\ba
\g_{\max}\equiv\max_{l \in C^{\perp}, \a}{\gamma_{\alpha}(\e_0-\e_l)} \ .
\label{eq:g_max}
\ea 
Using the spectral decomposition $\rho(0) = \sum_i \lambda_i \ketbra{i}$, it is clear that $\Tr[\rho(0)F_{\a}^{\dagger}\Pi_l F_{\a}] = \sum_i \lambda_i \| \Pi_l F_{\a} \ket{i}\|^2\geq 0$. Therefore, using Eq.~\eqref{eq:P_C}, the excitation rate satisfies the bound
\begin{align}
	 &|R|
	 \leq \g_{\max} \sum_{\alpha} \sum_{l \in C^{\perp}}\Tr[\rho(0)F_{\a}^{\dagger}\Pi_l F_{\a}] \notag \\ 
	 &=
	 \g_{\max} \left(\sum_{\alpha} \Tr[\rho(0)F_{\a}^{\dagger} F_{\a}]-\sum_{\alpha}\Tr[\rho(0)F_{\a}^{\dagger}P_C F_{\a}]\right) \notag \\
	 &\leq
	 \g_{\max} \sum_{\alpha}\Tr[\rho(0)F_{\a}^{\dagger} F_{\a}] 
\label{eq:poly-growth}
\end{align}
None of the terms in the last sum depends on the number of qubits $n$. The number of terms itself can increase at most polynomially in $n$, due to the sum over $\alpha$ [both the explicit one in Eq.~\eqref{eq:poly-growth} and also the implicit one in $F_{\a} =\sum_{\a'}(U^\dagger)_{\a\a'}A_{\a'}$]. This proves that the excitation rate grows at most polynomially in $n$.

\subsection{The excitation rate is exponentially suppressed by the energy penalty}
Next, let us show that for reasonable models of the bath the excitation rate is exponentially suppressed with increasing energy penalty $\eta_{\mrp}$. If the bath is in thermal equilibrium at inverse temperature $\b$, then under rather general conditions (analyticity of the bath correlation function in a strip) the matrix of decay rates satisfies the quantum detailed balance, or Kubo-Martin-Schwinger (KMS) condition \cite{KMS}: $\g_{\alpha \b}(-\o) = e^{-\b \o}\g_{\b \alpha}(\o)$.
The diagonalization used above then implies that the eigenvalues of the $\gamma$ matrix also satisfy the KMS condition, i.e., 
\begin{eqnarray}
\g_{\alpha}(-\o) = e^{-\b \o}\g_{\alpha}(\o) \ .
\label{eq:KMS}
\end{eqnarray}

Let $\Pi_l$ denote an eigenprojector of $H$ with energy $\epsilon_l = \Tr[\Pi_l H]$. It follows from Eq.~\eqref{eq:H(t)} and $[\overline{H}_S, H_{\mrp}]=0$ that these are simultaneous eigenstates of $\overline{H}_S$ and $H_{\mrp}$ as well. Let us assume that $H_{\mrp}$ has a ground state gap $g = \min_{l \in C^\perp}  \Tr[(\Pi_l -\Pi_0)H_{\mrp}]$. We have $\forall l \in C^{\perp}$:  
\begin{align}
\e_l -\e_0 &=\Tr[\Pi_l(\overline{H}_S +\eta_{\mrp} H_{\mrp})]-\Tr[\Pi_0(\overline{H}_S +\eta_{\mrp} H_{\mrp})]\notag \\
&=\Tr[(\Pi_l-\Pi_0)\overline{H}_S] +\eta_{\mrp} \Tr[(\Pi_l -\Pi_0)H_{\mrp}] \notag \\
&\geq \eta_{\mrp}g \ .
\end{align}
When $H_{\mrp}$ is a sum of commuting terms, as is true for the stabilizer construction we consider here, the gap $g$ is guaranteed to be a constant \cite{Bravyi:2003tx}.\footnote{For $H_{\mrp}$ that is a sum of non-commuting terms, e.g., when it is chosen as a sum of gauge group elements \cite{Jiang:2015kx,Marvian-Lidar:16}, $g$ may decrease with increasing system size. Even this case remains interesting if the gap of $\overline{H}_S$ decreases faster in the system size than $g$ \cite{Marvian-Lidar:16}.} 

Now, using the KMS condition~\eqref{eq:KMS}, we have:
\begin{align}
 \g_{\alpha}(\e_0 -\e_l) &= e^{-\b (\e_l -\e_0)}\g_{\alpha}(\e_l -\e_0) \notag \\
 &\leq e^{-\b g \eta_{\mrp}} \g_{\alpha}(\e_l -\e_0) \ . 
\end{align}
It follows that
\begin{align}
\g_{\max} \leq e^{-\b g \eta_{\mrp}} \max_{l \in C^{\perp}, \a}\gamma_{\alpha}(\e_l-\e_0)\ ,
\end{align}
and thus, the  bound on $|R|$ depends on $\max_{l \in C^{\perp}, \a}\gamma_{\alpha}(\e_l-\e_0) = \max_{l \in C^{\perp}, \a}\gamma_{\alpha}(\overline{\o}_l + \eta_{\mrp} \xi_l-\e_0)$, where we used Eq.~\eqref{eq:e_l}.
To ensure a non-trivial bound on $|R|$ this quantity has to be finite, which is a natural assumption. For example, for a bath satisfying an Ohmic-like relation of the form $\g(\o) = \mu \o^k e^{-\o/\o_c}$ for $\o>0$, where $\o_c$ is a finite cutoff frequency, the maximum value of $\g(\o)$ is $\mu (k\o_c)^k e^{-k}$. Even if this is not the case (e.g., in the quantum optical master equation $\g(\o) \propto \o^3$ for sufficiently large $\o$ \cite{Breuer:2002}) it is reasonable to assume that the system itself imposes a high-frequency cutoff, i.e., that $\max_{l \in C^{\perp}}\{\overline{\o}_l,\xi_l\}< \infty$.\footnote{This is certainly reasonable for condensed matter systems, where the finite number density naturally imposes a high-frequency cutoff, such as a Debye frequency.}

We also assume that $\g(\o)$ is a polynomial function (or any subexponential function in $\o$) for $\o>0$; this too is an assumption that is compatible with all commonly used bath models \cite{Breuer:2002}. 

Combining this with Eq.~\eqref{eq:poly-growth}, we have:
\begin{align}
|R| 
\leq \exp(-\b g \eta_{\mrp})\text{poly}(\eta_{\mrp}) \text{poly}(n)\ .
\label{eq:poly-growth-purity}
\end{align}
It follows that the excitation rate  is exponentially suppressed as the penalty strength $\eta_{\mrp}$ is increased. In other words, by using stabilizer error detecting codes (constant $g$), for Markovian models with a thermal bath that satisfy our assumptions above, to  keep the initial  excitation rate (or purity decay) out of the codespace constant while the system size $n$ increases, \emph{one only needs to increase the strength of energy penalty, $\eta_{\mrp}$, logarithmically in $n$}, at any fixed inverse temperature $\b$. The flatter the initial purity decay, the longer the adiabatic or holonomic quantum computation will proceed in the ground state.

\subsection{Relation to the JFS work}

In the pioneering JFS work \cite{jordan2006error}, a very similar result to Eq.~\eqref{eq:poly-growth-purity} was already established, under somewhat less general conditions.  Rather than dealing with a general Markovian master equation, they assumed a particular system of spins weakly coupled to a photon bath and a pure initial state. They then provided the lowest-weight possible subspace stabilizer codes for detecting $1$-local and $2$-local noise compatible with the error suppression scheme. Here, following and generalizing the JFS proof technique and providing all the necessary details, we generalized the suppression result to arbitrary Markovian master equation in Lindblad form and arbitrary stabilizer subspace error detection codes, while allowing for a degenerate initial state. We now proceed to establish the result even more generally, for Markovian master equations derived without the rotating wave approximation.

\section{Error suppression for non-Lindblad Markovian master equations}
\label{sec:3}

The derivation from first principles of a master equation with a dissipator in Lindblad form [Eq.~\eqref{eq:diss}] requires several approximations \cite{Breuer:2002}. Prominent among these is the rotating wave approximation (RWA), whose validity has often been questioned \cite{PhysRevLett.51.1108.3,PhysRevA.43.2430,Ford1997377,Schaller:2008aa,Fleming2010,Pekola:2010oj,PhysRevLett.108.033601,Majenz:2013qw}. Ref.~\cite{ABLZ:12-SI} presented a derivation not only of the Lindblad-form adiabatic master equation~\eqref{eq:ME}, which required the use of the RWA and guarantees complete positivity, but also a so-called double sided adiabatic master equation (DSAME), derived without the RWA (the Lindblad form follows from the latter master equation after using the RWA). In this section we show that the DSAME also exhibits the same suppression of purity decay or the excitation rate out of the ground space. Thus, the suppression effect does not depend on the RWA.

The DSAME has the following form:
\bes
\begin{align}
\dot{\rho}&=-i [H(t),\rho] + \tilde{D}[\rho]\\
\tilde{D}[\rho]&=\sum_{\a\b} \sum_{ll'} \G_{\a\b}(\o_{ll'}) [\Pi_lA_\b \Pi_{l'} \rho, A_\a] + \text{h.c.} \ , \notag \\
\label{eq:DSAME}
\end{align}
\ees
where h.c. denotes the Hermitian conjugate. Note the non-Lindblad form of the dissipator $\tilde{D}$ and the absence of an explicit Lamb shift term (such a term, i.e., a Hermitian part, can nevertheless be separated from $\tilde{D}$). Here the frequencies are the  time-dependent Bohr frequencies of the system: $\o_{ll'}(t) = \e_{l'}(t)-\e_l(t)$, where $H(t)\ket{\e_l(t)}=\e_l(t)\ket{\e_l(t)}$, and
\beq
\G_{\a\b}(\o) = \int_{0}^{\infty}dt\ e^{i\o t} \langle \mathcal{B}_{\a\b}(t)\rangle
\eeq 
is the one-sided Fourier transform of the bath correlation function. The double-sided and one-sided Fourier transforms are related via
\beq
\G_{\a\b}(\o) = \frac{1}{2}\g_{\a\b}(\o) + i S_{\a\b}(\o)\ ,
\label{eq:GgS}
\eeq
where $S_{\a\b}(\o) = S^*_{\b\a}(\o)$ is the remaining Cauchy principal value (see, e.g., ~Ref.~\cite{ABLZ:12-SI}).

We again calculate the excitation rate $\Tr[\Pi_0 \dot{\rho}(0)]$ for a state initialized in the ground subspace of the Hamiltonian $H$, with energy $\e_{0}$, i.e., $\rho(0)=\Pi_0 \rho(0) \Pi_0$. First:
\begin{align}
&\Tr\{\Pi_0\sum_{ll'} \G_{\a\b}(\o_{ll'}) [\Pi_lA_\b \Pi_{l'} \rho(0), A_\a]\} \notag\\
&= \sum_{ll'} \G_{\a\b}(\o_{ll'}) \left( \Tr\{\Pi_0 \Pi_lA_\b \Pi_{l'}[\Pi_0 \rho(0) \Pi_0] A_\a\} \right. \\
&\qquad\qquad\qquad\qquad \left. -\Tr\{\Pi_0 A_\a\Pi_lA_\b \Pi_{l'}[\Pi_0 \rho(0) \Pi_0]\}\right) \notag \\
&= \G_{\a\b}(0) \Tr[\rho(0) A_\a \Pi_0 A_\b] -\sum_{l} \G_{\a\b}(\o_{0l})\Tr[\rho(0) A_\a\Pi_lA_\b]\ . \notag
\end{align}
Next, after subtracting the $l=0$ term, we are left just with the sum over $l\neq 0$. Thus, using Eq.~\eqref{eq:GgS}:
\begin{align}
&\Tr\{\Pi_0\tilde{D}[\rho(0)]\} =\\
&- \sum_{\a\b} \sum_{l \neq 0} [\frac{1}{2}\g_{\a\b}(\o_{0l}) + i S_{\a\b}(\o_{0l})] \Tr[\rho(0) A_\a\Pi_lA_\b] +\text{h.c.}\notag
\end{align}
Accounting for the fact that without loss of generality we can always choose the system operators $A_\a$ to be Hermitian, and that the sum over all $\a$ and $\b$ allows us to interchange the order of summation, the imaginary part vanishes after summation with the Hermitian conjugate,
and we are left exactly with Eq.~\eqref{eq:rate} for the Lindblad form. The unitary part has no effect in the DSAME either (i.e., $\Tr\{\rho(0)[H(0),\rho(0)]\}=0$). Therefore, the same conclusions as reported in the previous section for master equations in Lindblad form, follow for the DSAME about the excitation rate out of the ground subspace of the Hamiltonian.

\section{Conclusion}
We have extended the JFS result \cite{jordan2006error}, that it suffices to increase the energy penalty logarithmically with system size in order to protect AQC against excitations out of the ground state, to general Markovian dynamics and mixed states. We have also pointed out that these results apply to HQC, and shown that the same results continue to hold even if the master equation is not in Lindblad form, i.e., without assuming the rotating wave approximation. These results only concern the initial excitation rate. A natural next generalization of these results is to subsystem codes and longer evolutions. 
  
\acknowledgments
We thank Tameem Albash for useful discussions. This work was supported under ARO Grant No. W911NF-12-1-0523, ARO MURI Grants No. W911NF-11-1-0268 and No. W911NF-15-1-0582, and NSF Grant No. INSPIRE-1551064.

\appendix


\section{Proof of Eq.~\eqref{eq:R}}
\label{app:A}

The excitation rate is $R=\Tr[\dot{\Pi}_0{\rho}(0)] + \Tr[{\Pi_0}\dot{\rho}(0)]$. Let us prove that the first term vanishes, which will prove the second equality in Eq.~\eqref{eq:R}. 

The initial state satisfies $\rho(0) = \Pi_0 \rho(0) \Pi_0$, so that $ \Tr[\dot\Pi_0 {\rho}(0)] =  \Tr[\Pi_0\dot\Pi_0 \Pi_0 {\rho}(0)]$. Now, differentiating the identity $\Pi_0^2 = \Pi_0$ yields
\begin{align}
\Pi_0 \dot{\Pi}_0 + \dot\Pi_0 \Pi_0= \dot\Pi_0\ &\Longrightarrow \ \dot\Pi_0 \Pi_0 = \Pi_0^\perp \dot\Pi_0 \notag\\
& \Longrightarrow \ \Pi_0\dot\Pi_0 \Pi_0 = 0 \ ,
\end{align}
where $\Pi_0^\perp = I-\Pi_0$.

\section{Proof of Eq.~\eqref{eq:rate}}
\label{app:B}

We explicitly compute the terms that need to be summed. We will use the fact that the initial state is in $\Pi_0$: $\rho(0)= \Pi_0 \rho(0) \Pi_0$. First:
\begin{align}
&\sum_\omega \gamma_{\alpha\beta}(\omega)\Tr[\Pi_0 A_{\beta}(\omega)\rho(0)A_{\alpha}^{\dagger}(\omega)]  \notag \\
&\quad = \sum_\omega \gamma_{\alpha\beta}(\omega)\Tr[\Pi_0 A_{\beta}(\omega)\Pi_0 \rho(0) \Pi_0 A_{\alpha}^{\dagger}(\omega)]  \notag \\
&\quad = \sum_\omega \gamma_{\alpha\beta}(\omega)\sum_{\epsilon_{l'}-\epsilon_l=\omega}\sum_{\epsilon_{l'''}-\epsilon_{l''}=\omega}\notag \\
&\qquad \Tr[\Pi_0({\Pi_lA_{\beta}\Pi_{l'}})\Pi_0 \rho(0) \Pi_0 (\Pi_{l'''}A^\dagger_{\alpha}\Pi_{l''})] \notag   \\
&\quad=\gamma_{\alpha\beta}(0)\Tr[\Pi_0A_{\b}\rho(0) A_{\alpha}^{\dagger}] \notag \\
&\quad=\gamma_{\alpha\beta}(0)\Tr[\rho(0)A_{\alpha}^{\dagger}\Pi_0A_{\b} ]
\end{align}

Second, similarly:
\begin{align}
& \sum_\omega \gamma_{\alpha\beta}(\omega) \Tr[\Pi_0A_{\alpha}^{\dagger}(\omega)A_{\beta}(\omega)\rho(0)]  \notag \\
& \quad =\sum_\omega \gamma_{\alpha\beta}(\omega) \sum_{\epsilon_{l'''}-\epsilon_{l''}=\omega}\sum_{\epsilon_{l'}-\epsilon_l=\omega}\notag \\
&\qquad \Tr[\Pi_0({\Pi_{l'''}A^\dagger_{\alpha}\Pi_{l''}})
({\Pi_lA_{\beta}\Pi_{l'}})\Pi_0 \rho(0)] \notag \\
& \quad = \sum_{l} \gamma_{\alpha\beta}(\e_0-\e_l) \Tr[\Pi_0{A^\dagger_{\alpha}\Pi_l}{A_{\beta} \rho(0)}] \notag \\
& \quad = \sum_{l} \gamma_{\alpha\beta}(\e_0-\e_l) \Tr[\rho(0){A^\dagger_{\alpha}\Pi_l}{A_{\beta} }]\ .
\end{align}
We also note that $\Tr[\Pi_0 A_{\alpha}^{\dagger}(\omega)A_{\beta}(\omega)\rho(0)] =\Tr[\Pi_0 \rho(0) A_{\alpha}^{\dagger}(\omega)A_{\beta}(\omega)] $, and so both terms of the anti-commutator produce the same result. Adding these terms according to the dissipator, Eq.~\eqref{eq:diss}, yields Eq.~\eqref{eq:rate}.

\bibliography{refs}

\begin{thebibliography}{61}%
\makeatletter
\providecommand \@ifxundefined [1]{%
 \@ifx{#1\undefined}
}%
\providecommand \@ifnum [1]{%
 \ifnum #1\expandafter \@firstoftwo
 \else \expandafter \@secondoftwo
 \fi
}%
\providecommand \@ifx [1]{%
 \ifx #1\expandafter \@firstoftwo
 \else \expandafter \@secondoftwo
 \fi
}%
\providecommand \natexlab [1]{#1}%
\providecommand \enquote  [1]{``#1''}%
\providecommand \bibnamefont  [1]{#1}%
\providecommand \bibfnamefont [1]{#1}%
\providecommand \citenamefont [1]{#1}%
\providecommand \href@noop [0]{\@secondoftwo}%
\providecommand \href [0]{\begingroup \@sanitize@url \@href}%
\providecommand \@href[1]{\@@startlink{#1}\@@href}%
\providecommand \@@href[1]{\endgroup#1\@@endlink}%
\providecommand \@sanitize@url [0]{\catcode `\\12\catcode `\$12\catcode
  `\&12\catcode `\#12\catcode `\^12\catcode `\_12\catcode `\%12\relax}%
\providecommand \@@startlink[1]{}%
\providecommand \@@endlink[0]{}%
\providecommand \url  [0]{\begingroup\@sanitize@url \@url }%
\providecommand \@url [1]{\endgroup\@href {#1}{\urlprefix }}%
\providecommand \urlprefix  [0]{URL }%
\providecommand \Eprint [0]{\href }%
\providecommand \doibase [0]{http://dx.doi.org/}%
\providecommand \selectlanguage [0]{\@gobble}%
\providecommand \bibinfo  [0]{\@secondoftwo}%
\providecommand \bibfield  [0]{\@secondoftwo}%
\providecommand \translation [1]{[#1]}%
\providecommand \BibitemOpen [0]{}%
\providecommand \bibitemStop [0]{}%
\providecommand \bibitemNoStop [0]{.\EOS\space}%
\providecommand \EOS [0]{\spacefactor3000\relax}%
\providecommand \BibitemShut  [1]{\csname bibitem#1\endcsname}%
\let\auto@bib@innerbib\@empty
\bibitem [{\citenamefont {Farhi}\ \emph {et~al.}(2000)\citenamefont {Farhi},
  \citenamefont {Goldstone}, \citenamefont {Gutmann},\ and\ \citenamefont
  {Sipser}}]{farhi_quantum_2000}%
  \BibitemOpen
  \bibfield  {author} {\bibinfo {author} {\bibfnamefont {E.}~\bibnamefont
  {Farhi}}, \bibinfo {author} {\bibfnamefont {J.}~\bibnamefont {Goldstone}},
  \bibinfo {author} {\bibfnamefont {S.}~\bibnamefont {Gutmann}}, \ and\
  \bibinfo {author} {\bibfnamefont {M.}~\bibnamefont {Sipser}},\ }\href
  {http://arxiv.org/abs/quant-ph/0001106} {\bibfield  {journal} {\bibinfo
  {journal} {arXiv:quant-ph/0001106}\ } (\bibinfo {year} {2000})}\BibitemShut
  {NoStop}%
\bibitem [{\citenamefont {Aharonov}\ \emph {et~al.}(2007)\citenamefont
  {Aharonov}, \citenamefont {van Dam}, \citenamefont {Kempe}, \citenamefont
  {Landau}, \citenamefont {Lloyd},\ and\ \citenamefont
  {Regev}}]{aharonov_adiabatic_2007}%
  \BibitemOpen
  \bibfield  {author} {\bibinfo {author} {\bibfnamefont {D.}~\bibnamefont
  {Aharonov}}, \bibinfo {author} {\bibfnamefont {W.}~\bibnamefont {van Dam}},
  \bibinfo {author} {\bibfnamefont {J.}~\bibnamefont {Kempe}}, \bibinfo
  {author} {\bibfnamefont {Z.}~\bibnamefont {Landau}}, \bibinfo {author}
  {\bibfnamefont {S.}~\bibnamefont {Lloyd}}, \ and\ \bibinfo {author}
  {\bibfnamefont {O.}~\bibnamefont {Regev}},\ }\href {\doibase
  10.1137/S0097539705447323} {\bibfield  {journal} {\bibinfo  {journal} {SIAM
  J. Comput.}\ }\textbf {\bibinfo {volume} {37}},\ \bibinfo {pages} {166}
  (\bibinfo {year} {2007})}\BibitemShut {NoStop}%
\bibitem [{\citenamefont {Oliveira}\ and\ \citenamefont
  {Terhal}(2008)}]{OliveiraGadget}%
  \BibitemOpen
  \bibfield  {author} {\bibinfo {author} {\bibfnamefont {R.}~\bibnamefont
  {Oliveira}}\ and\ \bibinfo {author} {\bibfnamefont {B.~M.}\ \bibnamefont
  {Terhal}},\ }\href {http://dl.acm.org/citation.cfm?id=2016985.2016987}
  {\bibfield  {journal} {\bibinfo  {journal} {Quantum Info. Comput.}\ }\textbf
  {\bibinfo {volume} {8}},\ \bibinfo {pages} {900} (\bibinfo {year}
  {2008})}\BibitemShut {NoStop}%
\bibitem [{\citenamefont {Mizel}\ \emph {et~al.}(2007)\citenamefont {Mizel},
  \citenamefont {Lidar},\ and\ \citenamefont
  {Mitchell}}]{PhysRevLett.99.070502}%
  \BibitemOpen
  \bibfield  {author} {\bibinfo {author} {\bibfnamefont {A.}~\bibnamefont
  {Mizel}}, \bibinfo {author} {\bibfnamefont {D.~A.}\ \bibnamefont {Lidar}}, \
  and\ \bibinfo {author} {\bibfnamefont {M.}~\bibnamefont {Mitchell}},\ }\href
  {\doibase 10.1103/PhysRevLett.99.070502} {\bibfield  {journal} {\bibinfo
  {journal} {Phys. Rev. Lett.}\ }\textbf {\bibinfo {volume} {99}},\ \bibinfo
  {pages} {070502} (\bibinfo {year} {2007})}\BibitemShut {NoStop}%
\bibitem [{\citenamefont {Breuckmann}\ and\ \citenamefont
  {Terhal}(2014)}]{Breuckmann:2014}%
  \BibitemOpen
  \bibfield  {author} {\bibinfo {author} {\bibfnamefont {N.~P.}\ \bibnamefont
  {Breuckmann}}\ and\ \bibinfo {author} {\bibfnamefont {B.~M.}\ \bibnamefont
  {Terhal}},\ }\href {http://stacks.iop.org/1751-8121/47/i=19/a=195304}
  {\bibfield  {journal} {\bibinfo  {journal} {Journal of Physics A:
  Mathematical and Theoretical}\ }\textbf {\bibinfo {volume} {47}},\ \bibinfo
  {pages} {195304} (\bibinfo {year} {2014})}\BibitemShut {NoStop}%
\bibitem [{\citenamefont {Gosset}\ \emph {et~al.}(2015)\citenamefont {Gosset},
  \citenamefont {Terhal},\ and\ \citenamefont {Vershynina}}]{Gosset:2014rp}%
  \BibitemOpen
  \bibfield  {author} {\bibinfo {author} {\bibfnamefont {D.}~\bibnamefont
  {Gosset}}, \bibinfo {author} {\bibfnamefont {B.~M.}\ \bibnamefont {Terhal}},
  \ and\ \bibinfo {author} {\bibfnamefont {A.}~\bibnamefont {Vershynina}},\
  }\href {http://link.aps.org/doi/10.1103/PhysRevLett.114.140501} {\bibfield
  {journal} {\bibinfo  {journal} {Physical Review Letters}\ }\textbf {\bibinfo
  {volume} {114}},\ \bibinfo {pages} {140501} (\bibinfo {year}
  {2015})}\BibitemShut {NoStop}%
\bibitem [{\citenamefont {Lloyd}\ and\ \citenamefont
  {Terhal}(2016)}]{Lloyd:2016}%
  \BibitemOpen
  \bibfield  {author} {\bibinfo {author} {\bibfnamefont {S.}~\bibnamefont
  {Lloyd}}\ and\ \bibinfo {author} {\bibfnamefont {B.~M.}\ \bibnamefont
  {Terhal}},\ }\href {http://stacks.iop.org/1367-2630/18/i=2/a=023042}
  {\bibfield  {journal} {\bibinfo  {journal} {New Journal of Physics}\ }\textbf
  {\bibinfo {volume} {18}},\ \bibinfo {pages} {023042} (\bibinfo {year}
  {2016})}\BibitemShut {NoStop}%
\bibitem [{\citenamefont {Albash}\ and\ \citenamefont
  {Lidar}(2016)}]{Albash-Lidar:RMP}%
  \BibitemOpen
  \bibfield  {author} {\bibinfo {author} {\bibfnamefont {T.}~\bibnamefont
  {Albash}}\ and\ \bibinfo {author} {\bibfnamefont {D.~A.}\ \bibnamefont
  {Lidar}},\ }\href {http://arXiv.org/abs/1611.04471} {\bibfield  {journal}
  {\bibinfo  {journal} {arXiv:1611.04471}\ } (\bibinfo {year}
  {2016})}\BibitemShut {NoStop}%
\bibitem [{\citenamefont {Jansen}\ \emph {et~al.}(2007)\citenamefont {Jansen},
  \citenamefont {Ruskai},\ and\ \citenamefont {Seiler}}]{Jansen:07}%
  \BibitemOpen
  \bibfield  {author} {\bibinfo {author} {\bibfnamefont {S.}~\bibnamefont
  {Jansen}}, \bibinfo {author} {\bibfnamefont {M.-B.}\ \bibnamefont {Ruskai}},
  \ and\ \bibinfo {author} {\bibfnamefont {R.}~\bibnamefont {Seiler}},\ }\href
  {http://scitation.aip.org/content/aip/journal/jmp/48/10/10.1063/1.2798382}
  {\bibfield  {journal} {\bibinfo  {journal} {J. Math. Phys.}\ }\textbf
  {\bibinfo {volume} {48}},\ \bibinfo {pages} {102111} (\bibinfo {year}
  {2007})}\BibitemShut {NoStop}%
\bibitem [{\citenamefont {Zanardi}\ and\ \citenamefont {Rasetti}(1999)}]{HQC}%
  \BibitemOpen
  \bibfield  {author} {\bibinfo {author} {\bibfnamefont {P.}~\bibnamefont
  {Zanardi}}\ and\ \bibinfo {author} {\bibfnamefont {M.}~\bibnamefont
  {Rasetti}},\ }\href {\doibase dx.doi.org/10.1016/S0375-9601(99)00803-8}
  {\bibfield  {journal} {\bibinfo  {journal} {Physics Letters A}\ }\textbf
  {\bibinfo {volume} {264}},\ \bibinfo {pages} {94} (\bibinfo {year}
  {1999})}\BibitemShut {NoStop}%
\bibitem [{\citenamefont {Duan}\ \emph {et~al.}(2001)\citenamefont {Duan},
  \citenamefont {Cirac},\ and\ \citenamefont {Zoller}}]{Duan:2001ff}%
  \BibitemOpen
  \bibfield  {author} {\bibinfo {author} {\bibfnamefont {L.~M.}\ \bibnamefont
  {Duan}}, \bibinfo {author} {\bibfnamefont {J.~I.}\ \bibnamefont {Cirac}}, \
  and\ \bibinfo {author} {\bibfnamefont {P.}~\bibnamefont {Zoller}},\ }\href
  {http://www.sciencemag.org/content/292/5522/1695.abstract N2 - We propose an
  experimentally feasible scheme to achieve quantum computation based solely on
  geometric manipulations of a quantum system. The desired geometric operations
  are obtained by driving the quantum system to undergo appropriate adiabatic
  cyclic evolutions. Our implementation of the all-geometric quantum
  computation is based on laser manipulation of a set of trapped ions. An
  all-geometric approach, apart from its fundamental interest, offers a
  possible method for robust quantum computation.} {\bibfield  {journal}
  {\bibinfo  {journal} {Science}\ }\textbf {\bibinfo {volume} {292}},\ \bibinfo
  {pages} {1695} (\bibinfo {year} {2001})}\BibitemShut {NoStop}%
\bibitem [{\citenamefont {Recati}\ \emph {et~al.}(2002)\citenamefont {Recati},
  \citenamefont {Calarco}, \citenamefont {Zanardi}, \citenamefont {Cirac},\
  and\ \citenamefont {Zoller}}]{Recati:02}%
  \BibitemOpen
  \bibfield  {author} {\bibinfo {author} {\bibfnamefont {A.}~\bibnamefont
  {Recati}}, \bibinfo {author} {\bibfnamefont {T.}~\bibnamefont {Calarco}},
  \bibinfo {author} {\bibfnamefont {P.}~\bibnamefont {Zanardi}}, \bibinfo
  {author} {\bibfnamefont {J.~I.}\ \bibnamefont {Cirac}}, \ and\ \bibinfo
  {author} {\bibfnamefont {P.}~\bibnamefont {Zoller}},\ }\href
  {http://link.aps.org/doi/10.1103/PhysRevA.66.032309} {\bibfield  {journal}
  {\bibinfo  {journal} {Physical Review A}\ }\textbf {\bibinfo {volume} {66}},\
  \bibinfo {pages} {032309} (\bibinfo {year} {2002})}\BibitemShut {NoStop}%
\bibitem [{\citenamefont {Lidar}\ and\ \citenamefont
  {Brun}(2013)}]{Lidar-Brun:book}%
  \BibitemOpen
  \bibinfo {editor} {\bibfnamefont {D.}~\bibnamefont {Lidar}}\ and\ \bibinfo
  {editor} {\bibfnamefont {T.}~\bibnamefont {Brun}},\ eds.,\ \href
  {http://www.cambridge.org/9780521897877} {\emph {\bibinfo {title} {Quantum
  Error Correction}}}\ (\bibinfo  {publisher} {Cambridge University Press},\
  \bibinfo {address} {{Cambridge, UK}},\ \bibinfo {year} {2013})\BibitemShut
  {NoStop}%
\bibitem [{\citenamefont {Childs}\ \emph {et~al.}(2001)\citenamefont {Childs},
  \citenamefont {Farhi},\ and\ \citenamefont
  {Preskill}}]{childs_robustness_2001}%
  \BibitemOpen
  \bibfield  {author} {\bibinfo {author} {\bibfnamefont {A.~M.}\ \bibnamefont
  {Childs}}, \bibinfo {author} {\bibfnamefont {E.}~\bibnamefont {Farhi}}, \
  and\ \bibinfo {author} {\bibfnamefont {J.}~\bibnamefont {Preskill}},\ }\href
  {\doibase 10.1103/PhysRevA.65.012322} {\bibfield  {journal} {\bibinfo
  {journal} {Phys. Rev. A}\ }\textbf {\bibinfo {volume} {65}},\ \bibinfo
  {pages} {012322} (\bibinfo {year} {2001})}\BibitemShut {NoStop}%
\bibitem [{\citenamefont {Sarandy}\ and\ \citenamefont
  {Lidar}(2005)}]{PhysRevLett.95.250503}%
  \BibitemOpen
  \bibfield  {author} {\bibinfo {author} {\bibfnamefont {M.~S.}\ \bibnamefont
  {Sarandy}}\ and\ \bibinfo {author} {\bibfnamefont {D.~A.}\ \bibnamefont
  {Lidar}},\ }\href {http://link.aps.org/doi/10.1103/PhysRevLett.95.250503}
  {\bibfield  {journal} {\bibinfo  {journal} {Phys. Rev. Lett.}\ }\textbf
  {\bibinfo {volume} {95}},\ \bibinfo {pages} {250503} (\bibinfo {year}
  {2005})}\BibitemShut {NoStop}%
\bibitem [{\citenamefont {Aberg}\ \emph {et~al.}(2005)\citenamefont {Aberg},
  \citenamefont {Kult},\ and\ \citenamefont {Sj\"oqvist}}]{Aberg:2005rt}%
  \BibitemOpen
  \bibfield  {author} {\bibinfo {author} {\bibfnamefont {J.}~\bibnamefont
  {Aberg}}, \bibinfo {author} {\bibfnamefont {D.}~\bibnamefont {Kult}}, \ and\
  \bibinfo {author} {\bibfnamefont {E.}~\bibnamefont {Sj\"oqvist}},\ }\href
  {http://link.aps.org/doi/10.1103/PhysRevA.72.042317} {\bibfield  {journal}
  {\bibinfo  {journal} {Phys. Rev. A}\ }\textbf {\bibinfo {volume} {72}},\
  \bibinfo {pages} {042317} (\bibinfo {year} {2005})}\BibitemShut {NoStop}%
\bibitem [{\citenamefont {{S. Ashhab, J. R. Johansson, and F.
  Nori}}(2006)}]{ashhab:052330}%
  \BibitemOpen
  \bibfield  {author} {\bibinfo {author} {\bibnamefont {{S. Ashhab, J. R.
  Johansson, and F. Nori}}},\ }\href {\doibase 10.1103/PhysRevA.74.052330}
  {\bibfield  {journal} {\bibinfo  {journal} {Phys. Rev. A}\ }\textbf {\bibinfo
  {volume} {74}},\ \bibinfo {eid} {052330} (\bibinfo {year}
  {2006})}\BibitemShut {NoStop}%
\bibitem [{\citenamefont {Tiersch}\ and\ \citenamefont
  {Sch{\"{u}}tzhold}(2007)}]{PhysRevA.75.062313}%
  \BibitemOpen
  \bibfield  {author} {\bibinfo {author} {\bibfnamefont {M.}~\bibnamefont
  {Tiersch}}\ and\ \bibinfo {author} {\bibfnamefont {R.}~\bibnamefont
  {Sch{\"{u}}tzhold}},\ }\href {\doibase 10.1103/PhysRevA.75.062313} {\bibfield
   {journal} {\bibinfo  {journal} {Phys. Rev. A}\ }\textbf {\bibinfo {volume}
  {75}},\ \bibinfo {pages} {062313} (\bibinfo {year} {2007})}\BibitemShut
  {NoStop}%
\bibitem [{\citenamefont {Amin}\ \emph
  {et~al.}(2009{\natexlab{a}})\citenamefont {Amin}, \citenamefont {Averin},\
  and\ \citenamefont {Nesteroff}}]{amin_decoherence_2009}%
  \BibitemOpen
  \bibfield  {author} {\bibinfo {author} {\bibfnamefont {M.~H.~S.}\
  \bibnamefont {Amin}}, \bibinfo {author} {\bibfnamefont {D.~V.}\ \bibnamefont
  {Averin}}, \ and\ \bibinfo {author} {\bibfnamefont {J.~A.}\ \bibnamefont
  {Nesteroff}},\ }\href {\doibase 10.1103/PhysRevA.79.022107} {\bibfield
  {journal} {\bibinfo  {journal} {Phys. Rev. A}\ }\textbf {\bibinfo {volume}
  {79}},\ \bibinfo {pages} {022107} (\bibinfo {year}
  {2009}{\natexlab{a}})}\BibitemShut {NoStop}%
\bibitem [{\citenamefont {Amin}\ \emph
  {et~al.}(2009{\natexlab{b}})\citenamefont {Amin}, \citenamefont {Truncik},\
  and\ \citenamefont {Averin}}]{PhysRevA.80.022303}%
  \BibitemOpen
  \bibfield  {author} {\bibinfo {author} {\bibfnamefont {M.~H.~S.}\
  \bibnamefont {Amin}}, \bibinfo {author} {\bibfnamefont {C.~J.~S.}\
  \bibnamefont {Truncik}}, \ and\ \bibinfo {author} {\bibfnamefont {D.~V.}\
  \bibnamefont {Averin}},\ }\href {\doibase 10.1103/PhysRevA.80.022303}
  {\bibfield  {journal} {\bibinfo  {journal} {Phys. Rev. A}\ }\textbf {\bibinfo
  {volume} {80}},\ \bibinfo {pages} {022303} (\bibinfo {year}
  {2009}{\natexlab{b}})}\BibitemShut {NoStop}%
\bibitem [{\citenamefont {Sarovar}\ and\ \citenamefont
  {Young}(2013)}]{Sarovar:2013kx}%
  \BibitemOpen
  \bibfield  {author} {\bibinfo {author} {\bibfnamefont {M.}~\bibnamefont
  {Sarovar}}\ and\ \bibinfo {author} {\bibfnamefont {K.~C.}\ \bibnamefont
  {Young}},\ }\href {http://stacks.iop.org/1367-2630/15/i=12/a=125032}
  {\bibfield  {journal} {\bibinfo  {journal} {New J. of Phys.}\ }\textbf
  {\bibinfo {volume} {15}},\ \bibinfo {pages} {125032} (\bibinfo {year}
  {2013})}\BibitemShut {NoStop}%
\bibitem [{\citenamefont {Albash}\ and\ \citenamefont
  {Lidar}(2015)}]{Albash:2015nx}%
  \BibitemOpen
  \bibfield  {author} {\bibinfo {author} {\bibfnamefont {T.}~\bibnamefont
  {Albash}}\ and\ \bibinfo {author} {\bibfnamefont {D.~A.}\ \bibnamefont
  {Lidar}},\ }\href {http://link.aps.org/doi/10.1103/PhysRevA.91.062320}
  {\bibfield  {journal} {\bibinfo  {journal} {Phys. Rev. A}\ }\textbf {\bibinfo
  {volume} {91}},\ \bibinfo {pages} {062320} (\bibinfo {year}
  {2015})}\BibitemShut {NoStop}%
\bibitem [{\citenamefont {Jordan}\ \emph {et~al.}(2006)\citenamefont {Jordan},
  \citenamefont {Farhi},\ and\ \citenamefont {Shor}}]{jordan2006error}%
  \BibitemOpen
  \bibfield  {author} {\bibinfo {author} {\bibfnamefont {S.~P.}\ \bibnamefont
  {Jordan}}, \bibinfo {author} {\bibfnamefont {E.}~\bibnamefont {Farhi}}, \
  and\ \bibinfo {author} {\bibfnamefont {P.~W.}\ \bibnamefont {Shor}},\ }\href
  {http://link.aps.org/doi/10.1103/PhysRevA.74.052322} {\bibfield  {journal}
  {\bibinfo  {journal} {{Phys. Rev. A}}\ }\textbf {\bibinfo {volume} {74}},\
  \bibinfo {pages} {052322} (\bibinfo {year} {2006})}\BibitemShut {NoStop}%
\bibitem [{\citenamefont {Bookatz}\ \emph {et~al.}(2015)\citenamefont
  {Bookatz}, \citenamefont {Farhi},\ and\ \citenamefont
  {Zhou}}]{Bookatz:2014uq}%
  \BibitemOpen
  \bibfield  {author} {\bibinfo {author} {\bibfnamefont {A.~D.}\ \bibnamefont
  {Bookatz}}, \bibinfo {author} {\bibfnamefont {E.}~\bibnamefont {Farhi}}, \
  and\ \bibinfo {author} {\bibfnamefont {L.}~\bibnamefont {Zhou}},\ }\href
  {http://link.aps.org/doi/10.1103/PhysRevA.92.022317} {\bibfield  {journal}
  {\bibinfo  {journal} {Physical Review A}\ }\textbf {\bibinfo {volume} {92}},\
  \bibinfo {pages} {022317} (\bibinfo {year} {2015})}\BibitemShut {NoStop}%
\bibitem [{\citenamefont {Marvian}(2016)}]{Marvian:2016aa}%
  \BibitemOpen
  \bibfield  {author} {\bibinfo {author} {\bibfnamefont {I.}~\bibnamefont
  {Marvian}},\ }\href {http://arXiv.org/abs/1602.03251} {\bibfield  {journal}
  {\bibinfo  {journal} {arXiv:1602.03251}\ } (\bibinfo {year}
  {2016})}\BibitemShut {NoStop}%
\bibitem [{\citenamefont {Jiang}\ and\ \citenamefont
  {Rieffel}(2015)}]{Jiang:2015kx}%
  \BibitemOpen
  \bibfield  {author} {\bibinfo {author} {\bibfnamefont {Z.}~\bibnamefont
  {Jiang}}\ and\ \bibinfo {author} {\bibfnamefont {E.~G.}\ \bibnamefont
  {Rieffel}},\ }\href {http://arXiv.org/abs/1511.01997} {\bibfield  {journal}
  {\bibinfo  {journal} {arXiv:1511.01997}\ } (\bibinfo {year}
  {2015})}\BibitemShut {NoStop}%
\bibitem [{\citenamefont {Marvian}\ and\ \citenamefont
  {Lidar}(2017)}]{Marvian-Lidar:16}%
  \BibitemOpen
  \bibfield  {author} {\bibinfo {author} {\bibfnamefont {M.}~\bibnamefont
  {Marvian}}\ and\ \bibinfo {author} {\bibfnamefont {D.~A.}\ \bibnamefont
  {Lidar}},\ }\href {\doibase 10.1103/PhysRevLett.118.030504} {\bibfield
  {journal} {\bibinfo  {journal} {Phys. Rev. Lett.}\ }\textbf {\bibinfo
  {volume} {118}},\ \bibinfo {pages} {030504} (\bibinfo {year}
  {2017})}\BibitemShut {NoStop}%
\bibitem [{\citenamefont {Johnson}\ \emph {et~al.}(2011)\citenamefont
  {Johnson}, \citenamefont {Amin}, \citenamefont {Gildert}, \citenamefont
  {Lanting}, \citenamefont {Hamze}, \citenamefont {Dickson}, \citenamefont
  {Harris}, \citenamefont {Berkley}, \citenamefont {Johansson}, \citenamefont
  {Bunyk}, \citenamefont {Chapple}, \citenamefont {Enderud}, \citenamefont
  {Hilton}, \citenamefont {Karimi}, \citenamefont {Ladizinsky}, \citenamefont
  {Ladizinsky}, \citenamefont {Oh}, \citenamefont {Perminov}, \citenamefont
  {Rich}, \citenamefont {Thom}, \citenamefont {Tolkacheva}, \citenamefont
  {Truncik}, \citenamefont {Uchaikin}, \citenamefont {Wang}, \citenamefont
  {Wilson},\ and\ \citenamefont {Rose}}]{DWave}%
  \BibitemOpen
  \bibfield  {author} {\bibinfo {author} {\bibfnamefont {M.~W.}\ \bibnamefont
  {Johnson}}, \bibinfo {author} {\bibfnamefont {M.~H.~S.}\ \bibnamefont
  {Amin}}, \bibinfo {author} {\bibfnamefont {S.}~\bibnamefont {Gildert}},
  \bibinfo {author} {\bibfnamefont {T.}~\bibnamefont {Lanting}}, \bibinfo
  {author} {\bibfnamefont {F.}~\bibnamefont {Hamze}}, \bibinfo {author}
  {\bibfnamefont {N.}~\bibnamefont {Dickson}}, \bibinfo {author} {\bibfnamefont
  {R.}~\bibnamefont {Harris}}, \bibinfo {author} {\bibfnamefont {A.~J.}\
  \bibnamefont {Berkley}}, \bibinfo {author} {\bibfnamefont {J.}~\bibnamefont
  {Johansson}}, \bibinfo {author} {\bibfnamefont {P.}~\bibnamefont {Bunyk}},
  \bibinfo {author} {\bibfnamefont {E.~M.}\ \bibnamefont {Chapple}}, \bibinfo
  {author} {\bibfnamefont {C.}~\bibnamefont {Enderud}}, \bibinfo {author}
  {\bibfnamefont {J.~P.}\ \bibnamefont {Hilton}}, \bibinfo {author}
  {\bibfnamefont {K.}~\bibnamefont {Karimi}}, \bibinfo {author} {\bibfnamefont
  {E.}~\bibnamefont {Ladizinsky}}, \bibinfo {author} {\bibfnamefont
  {N.}~\bibnamefont {Ladizinsky}}, \bibinfo {author} {\bibfnamefont
  {T.}~\bibnamefont {Oh}}, \bibinfo {author} {\bibfnamefont {I.}~\bibnamefont
  {Perminov}}, \bibinfo {author} {\bibfnamefont {C.}~\bibnamefont {Rich}},
  \bibinfo {author} {\bibfnamefont {M.~C.}\ \bibnamefont {Thom}}, \bibinfo
  {author} {\bibfnamefont {E.}~\bibnamefont {Tolkacheva}}, \bibinfo {author}
  {\bibfnamefont {C.~J.~S.}\ \bibnamefont {Truncik}}, \bibinfo {author}
  {\bibfnamefont {S.}~\bibnamefont {Uchaikin}}, \bibinfo {author}
  {\bibfnamefont {J.}~\bibnamefont {Wang}}, \bibinfo {author} {\bibfnamefont
  {B.}~\bibnamefont {Wilson}}, \ and\ \bibinfo {author} {\bibfnamefont
  {G.}~\bibnamefont {Rose}},\ }\href {\doibase 10.1038/nature10012} {\bibfield
  {journal} {\bibinfo  {journal} {Nature}\ }\textbf {\bibinfo {volume} {473}},\
  \bibinfo {pages} {194} (\bibinfo {year} {2011})}\BibitemShut {NoStop}%
\bibitem [{\citenamefont {Pudenz}\ \emph {et~al.}(2014)\citenamefont {Pudenz},
  \citenamefont {Albash},\ and\ \citenamefont {Lidar}}]{PAL:13}%
  \BibitemOpen
  \bibfield  {author} {\bibinfo {author} {\bibfnamefont {K.~L.}\ \bibnamefont
  {Pudenz}}, \bibinfo {author} {\bibfnamefont {T.}~\bibnamefont {Albash}}, \
  and\ \bibinfo {author} {\bibfnamefont {D.~A.}\ \bibnamefont {Lidar}},\ }\href
  {\doibase 10.1038/ncomms4243} {\bibfield  {journal} {\bibinfo  {journal}
  {Nat. Commun.}\ }\textbf {\bibinfo {volume} {5}},\ \bibinfo {pages} {3243}
  (\bibinfo {year} {2014})}\BibitemShut {NoStop}%
\bibitem [{\citenamefont {Matsuura}\ \emph {et~al.}(2016)\citenamefont
  {Matsuura}, \citenamefont {Nishimori}, \citenamefont {Albash},\ and\
  \citenamefont {Lidar}}]{MNAL:15}%
  \BibitemOpen
  \bibfield  {author} {\bibinfo {author} {\bibfnamefont {S.}~\bibnamefont
  {Matsuura}}, \bibinfo {author} {\bibfnamefont {H.}~\bibnamefont {Nishimori}},
  \bibinfo {author} {\bibfnamefont {T.}~\bibnamefont {Albash}}, \ and\ \bibinfo
  {author} {\bibfnamefont {D.~A.}\ \bibnamefont {Lidar}},\ }\href
  {http://link.aps.org/doi/10.1103/PhysRevLett.116.220501} {\bibfield
  {journal} {\bibinfo  {journal} {Physical Review Letters}\ }\textbf {\bibinfo
  {volume} {116}},\ \bibinfo {pages} {220501} (\bibinfo {year}
  {2016})}\BibitemShut {NoStop}%
\bibitem [{\citenamefont {Vinci}\ \emph {et~al.}(2016)\citenamefont {Vinci},
  \citenamefont {Albash},\ and\ \citenamefont {Lidar}}]{vinci2015nested}%
  \BibitemOpen
  \bibfield  {author} {\bibinfo {author} {\bibfnamefont {W.}~\bibnamefont
  {Vinci}}, \bibinfo {author} {\bibfnamefont {T.}~\bibnamefont {Albash}}, \
  and\ \bibinfo {author} {\bibfnamefont {D.~A.}\ \bibnamefont {Lidar}},\ }\href
  {http://dx.doi.org/10.1038/npjqi.2016.17} {\bibfield  {journal} {\bibinfo
  {journal} {Nature Quantum Information}\ }\textbf {\bibinfo {volume} {2}},\
  \bibinfo {pages} {16017} (\bibinfo {year} {2016})}\BibitemShut {NoStop}%
\bibitem [{\citenamefont {Solinas}\ \emph {et~al.}(2004)\citenamefont
  {Solinas}, \citenamefont {Zanardi},\ and\ \citenamefont
  {Zangh{\`\i}}}]{Solinas:04}%
  \BibitemOpen
  \bibfield  {author} {\bibinfo {author} {\bibfnamefont {P.}~\bibnamefont
  {Solinas}}, \bibinfo {author} {\bibfnamefont {P.}~\bibnamefont {Zanardi}}, \
  and\ \bibinfo {author} {\bibfnamefont {N.}~\bibnamefont {Zangh{\`\i}}},\
  }\href {http://link.aps.org/doi/10.1103/PhysRevA.70.042316} {\bibfield
  {journal} {\bibinfo  {journal} {Physical Review A}\ }\textbf {\bibinfo
  {volume} {70}},\ \bibinfo {pages} {042316} (\bibinfo {year}
  {2004})}\BibitemShut {NoStop}%
\bibitem [{\citenamefont {Wu}\ \emph {et~al.}(2005)\citenamefont {Wu},
  \citenamefont {Zanardi},\ and\ \citenamefont {Lidar}}]{Wu:2005aa}%
  \BibitemOpen
  \bibfield  {author} {\bibinfo {author} {\bibfnamefont {L.~A.}\ \bibnamefont
  {Wu}}, \bibinfo {author} {\bibfnamefont {P.}~\bibnamefont {Zanardi}}, \ and\
  \bibinfo {author} {\bibfnamefont {D.~A.}\ \bibnamefont {Lidar}},\ }\href
  {http://link.aps.org/doi/10.1103/PhysRevLett.95.130501} {\bibfield  {journal}
  {\bibinfo  {journal} {Physical Review Letters}\ }\textbf {\bibinfo {volume}
  {95}},\ \bibinfo {pages} {130501} (\bibinfo {year} {2005})}\BibitemShut
  {NoStop}%
\bibitem [{\citenamefont {Florio}\ \emph {et~al.}(2006)\citenamefont {Florio},
  \citenamefont {Facchi}, \citenamefont {Fazio}, \citenamefont {Giovannetti},\
  and\ \citenamefont {Pascazio}}]{Florio:06}%
  \BibitemOpen
  \bibfield  {author} {\bibinfo {author} {\bibfnamefont {G.}~\bibnamefont
  {Florio}}, \bibinfo {author} {\bibfnamefont {P.}~\bibnamefont {Facchi}},
  \bibinfo {author} {\bibfnamefont {R.}~\bibnamefont {Fazio}}, \bibinfo
  {author} {\bibfnamefont {V.}~\bibnamefont {Giovannetti}}, \ and\ \bibinfo
  {author} {\bibfnamefont {S.}~\bibnamefont {Pascazio}},\ }\href
  {http://link.aps.org/doi/10.1103/PhysRevA.73.022327} {\bibfield  {journal}
  {\bibinfo  {journal} {Physical Review A}\ }\textbf {\bibinfo {volume} {73}},\
  \bibinfo {pages} {022327} (\bibinfo {year} {2006})}\BibitemShut {NoStop}%
\bibitem [{\citenamefont {Sarandy}\ and\ \citenamefont
  {Lidar}(2006)}]{sarandy_abelian_2006}%
  \BibitemOpen
  \bibfield  {author} {\bibinfo {author} {\bibfnamefont {M.~S.}\ \bibnamefont
  {Sarandy}}\ and\ \bibinfo {author} {\bibfnamefont {D.~A.}\ \bibnamefont
  {Lidar}},\ }\href {\doibase 10.1103/PhysRevA.73.062101} {\bibfield  {journal}
  {\bibinfo  {journal} {Phys. Rev. A}\ }\textbf {\bibinfo {volume} {73}},\
  \bibinfo {pages} {062101} (\bibinfo {year} {2006})}\BibitemShut {NoStop}%
\bibitem [{\citenamefont {Parodi}\ \emph {et~al.}(2007)\citenamefont {Parodi},
  \citenamefont {Sassetti}, \citenamefont {Solinas},\ and\ \citenamefont
  {Zangh{\`\i}}}]{sol07}%
  \BibitemOpen
  \bibfield  {author} {\bibinfo {author} {\bibfnamefont {D.}~\bibnamefont
  {Parodi}}, \bibinfo {author} {\bibfnamefont {M.}~\bibnamefont {Sassetti}},
  \bibinfo {author} {\bibfnamefont {P.}~\bibnamefont {Solinas}}, \ and\
  \bibinfo {author} {\bibfnamefont {N.}~\bibnamefont {Zangh{\`\i}}},\ }\href
  {http://link.aps.org/doi/10.1103/PhysRevA.76.012337} {\bibfield  {journal}
  {\bibinfo  {journal} {Physical Review A}\ }\textbf {\bibinfo {volume} {76}},\
  \bibinfo {pages} {012337} (\bibinfo {year} {2007})}\BibitemShut {NoStop}%
\bibitem [{\citenamefont {Oreshkov}\ and\ \citenamefont
  {Calsamiglia}(2010)}]{oreshkov_adiabatic_2010}%
  \BibitemOpen
  \bibfield  {author} {\bibinfo {author} {\bibfnamefont {O.}~\bibnamefont
  {Oreshkov}}\ and\ \bibinfo {author} {\bibfnamefont {J.}~\bibnamefont
  {Calsamiglia}},\ }\href {\doibase 10.1103/PhysRevLett.105.050503} {\bibfield
  {journal} {\bibinfo  {journal} {Phys. Rev. Lett.}\ }\textbf {\bibinfo
  {volume} {105}},\ \bibinfo {pages} {050503} (\bibinfo {year}
  {2010})}\BibitemShut {NoStop}%
\bibitem [{\citenamefont {Xu}\ \emph {et~al.}(2012)\citenamefont {Xu},
  \citenamefont {Zhang}, \citenamefont {Tong}, \citenamefont {Sj{\"o}qvist},\
  and\ \citenamefont {Kwek}}]{Xu:2012aa}%
  \BibitemOpen
  \bibfield  {author} {\bibinfo {author} {\bibfnamefont {G.~F.}\ \bibnamefont
  {Xu}}, \bibinfo {author} {\bibfnamefont {J.}~\bibnamefont {Zhang}}, \bibinfo
  {author} {\bibfnamefont {D.~M.}\ \bibnamefont {Tong}}, \bibinfo {author}
  {\bibfnamefont {E.}~\bibnamefont {Sj{\"o}qvist}}, \ and\ \bibinfo {author}
  {\bibfnamefont {L.~C.}\ \bibnamefont {Kwek}},\ }\href
  {http://link.aps.org/doi/10.1103/PhysRevLett.109.170501} {\bibfield
  {journal} {\bibinfo  {journal} {Physical Review Letters}\ }\textbf {\bibinfo
  {volume} {109}},\ \bibinfo {pages} {170501} (\bibinfo {year}
  {2012})}\BibitemShut {NoStop}%
\bibitem [{\citenamefont {Oreshkov}\ \emph
  {et~al.}(2009{\natexlab{a}})\citenamefont {Oreshkov}, \citenamefont {Brun},\
  and\ \citenamefont {Lidar}}]{Oreshkov:2009bl}%
  \BibitemOpen
  \bibfield  {author} {\bibinfo {author} {\bibfnamefont {O.}~\bibnamefont
  {Oreshkov}}, \bibinfo {author} {\bibfnamefont {T.~A.}\ \bibnamefont {Brun}},
  \ and\ \bibinfo {author} {\bibfnamefont {D.~A.}\ \bibnamefont {Lidar}},\
  }\href {http://link.aps.org/doi/10.1103/PhysRevLett.102.070502} {\bibfield
  {journal} {\bibinfo  {journal} {Phys. Rev. Lett.}\ }\textbf {\bibinfo
  {volume} {102}},\ \bibinfo {pages} {070502} (\bibinfo {year}
  {2009}{\natexlab{a}})}\BibitemShut {NoStop}%
\bibitem [{\citenamefont {Oreshkov}\ \emph
  {et~al.}(2009{\natexlab{b}})\citenamefont {Oreshkov}, \citenamefont {Brun},\
  and\ \citenamefont {Lidar}}]{Oreshkov:2009lq}%
  \BibitemOpen
  \bibfield  {author} {\bibinfo {author} {\bibfnamefont {O.}~\bibnamefont
  {Oreshkov}}, \bibinfo {author} {\bibfnamefont {T.~A.}\ \bibnamefont {Brun}},
  \ and\ \bibinfo {author} {\bibfnamefont {D.~A.}\ \bibnamefont {Lidar}},\
  }\href {http://link.aps.org/doi/10.1103/PhysRevA.80.022325} {\bibfield
  {journal} {\bibinfo  {journal} {Phys. Rev. A}\ }\textbf {\bibinfo {volume}
  {80}},\ \bibinfo {pages} {022325} (\bibinfo {year}
  {2009}{\natexlab{b}})}\BibitemShut {NoStop}%
\bibitem [{\citenamefont {Albash}\ \emph {et~al.}(2012)\citenamefont {Albash},
  \citenamefont {Boixo}, \citenamefont {Lidar},\ and\ \citenamefont
  {Zanardi}}]{ABLZ:12-SI}%
  \BibitemOpen
  \bibfield  {author} {\bibinfo {author} {\bibfnamefont {T.}~\bibnamefont
  {Albash}}, \bibinfo {author} {\bibfnamefont {S.}~\bibnamefont {Boixo}},
  \bibinfo {author} {\bibfnamefont {D.~A.}\ \bibnamefont {Lidar}}, \ and\
  \bibinfo {author} {\bibfnamefont {P.}~\bibnamefont {Zanardi}},\ }\href
  {\doibase 10.1088/1367-2630/14/12/123016} {\bibfield  {journal} {\bibinfo
  {journal} {New J. of Phys.}\ }\textbf {\bibinfo {volume} {14}},\ \bibinfo
  {pages} {123016} (\bibinfo {year} {2012})}\BibitemShut {NoStop}%
\bibitem [{\citenamefont {Breuer}\ and\ \citenamefont
  {Petruccione}(2002)}]{Breuer:2002}%
  \BibitemOpen
  \bibfield  {author} {\bibinfo {author} {\bibfnamefont {H.-P.}\ \bibnamefont
  {Breuer}}\ and\ \bibinfo {author} {\bibfnamefont {F.}~\bibnamefont
  {Petruccione}},\ }\href@noop {} {\emph {\bibinfo {title} {The Theory of Open
  Quantum Systems}}}\ (\bibinfo  {publisher} {Oxford University Press},\
  \bibinfo {year} {2002})\BibitemShut {NoStop}%
\bibitem [{\citenamefont {{R. Alicki and K. Lendi}}(1987)}]{Alicki:87}%
  \BibitemOpen
  \bibfield  {author} {\bibinfo {author} {\bibnamefont {{R. Alicki and K.
  Lendi}}},\ }\href {http://www.springer.com/us/book/9783540708605} {\emph
  {\bibinfo {title} {{Quantum Dynamical Semigroups and Applications}}}},\
  \bibinfo {series} {{Lecture Notes in Physics}}, Vol.\ \bibinfo {volume}
  {286}\ (\bibinfo  {publisher} {{Springer-Verlag}},\ \bibinfo {address}
  {Berlin},\ \bibinfo {year} {1987})\BibitemShut {NoStop}%
\bibitem [{\citenamefont {Lidar}(2008)}]{PhysRevLett.100.160506}%
  \BibitemOpen
  \bibfield  {author} {\bibinfo {author} {\bibfnamefont {D.~A.}\ \bibnamefont
  {Lidar}},\ }\href {http://link.aps.org/doi/10.1103/PhysRevLett.100.160506}
  {\bibfield  {journal} {\bibinfo  {journal} {{Phys.~Rev.~Lett.}}\ }\textbf
  {\bibinfo {volume} {100}},\ \bibinfo {pages} {160506} (\bibinfo {year}
  {2008})}\BibitemShut {NoStop}%
\bibitem [{\citenamefont {Quiroz}\ and\ \citenamefont
  {Lidar}(2012)}]{PhysRevA.86.042333}%
  \BibitemOpen
  \bibfield  {author} {\bibinfo {author} {\bibfnamefont {G.}~\bibnamefont
  {Quiroz}}\ and\ \bibinfo {author} {\bibfnamefont {D.~A.}\ \bibnamefont
  {Lidar}},\ }\href {\doibase 10.1103/PhysRevA.86.042333} {\bibfield  {journal}
  {\bibinfo  {journal} {Phys. Rev. A}\ }\textbf {\bibinfo {volume} {86}},\
  \bibinfo {pages} {042333} (\bibinfo {year} {2012})}\BibitemShut {NoStop}%
\bibitem [{\citenamefont {Ganti}\ \emph {et~al.}(2014)\citenamefont {Ganti},
  \citenamefont {Onunkwo},\ and\ \citenamefont {Young}}]{Ganti:13}%
  \BibitemOpen
  \bibfield  {author} {\bibinfo {author} {\bibfnamefont {A.}~\bibnamefont
  {Ganti}}, \bibinfo {author} {\bibfnamefont {U.}~\bibnamefont {Onunkwo}}, \
  and\ \bibinfo {author} {\bibfnamefont {K.}~\bibnamefont {Young}},\ }\href
  {http://link.aps.org/doi/10.1103/PhysRevA.89.042313} {\bibfield  {journal}
  {\bibinfo  {journal} {Phys. Rev. A}\ }\textbf {\bibinfo {volume} {89}},\
  \bibinfo {pages} {042313} (\bibinfo {year} {2014})}\BibitemShut {NoStop}%
\bibitem [{\citenamefont {Paz-Silva}\ \emph {et~al.}(2012)\citenamefont
  {Paz-Silva}, \citenamefont {Rezakhani}, \citenamefont {Dominy},\ and\
  \citenamefont {Lidar}}]{PhysRevLett.108.080501}%
  \BibitemOpen
  \bibfield  {author} {\bibinfo {author} {\bibfnamefont {G.~A.}\ \bibnamefont
  {Paz-Silva}}, \bibinfo {author} {\bibfnamefont {A.~T.}\ \bibnamefont
  {Rezakhani}}, \bibinfo {author} {\bibfnamefont {J.~M.}\ \bibnamefont
  {Dominy}}, \ and\ \bibinfo {author} {\bibfnamefont {D.~A.}\ \bibnamefont
  {Lidar}},\ }\href {\doibase 10.1103/PhysRevLett.108.080501} {\bibfield
  {journal} {\bibinfo  {journal} {Phys. Rev. Lett.}\ }\textbf {\bibinfo
  {volume} {108}},\ \bibinfo {pages} {080501} (\bibinfo {year}
  {2012})}\BibitemShut {NoStop}%
\bibitem [{\citenamefont {Young}\ \emph {et~al.}(2013)\citenamefont {Young},
  \citenamefont {Sarovar},\ and\ \citenamefont {Blume-Kohout}}]{Young:13}%
  \BibitemOpen
  \bibfield  {author} {\bibinfo {author} {\bibfnamefont {K.~C.}\ \bibnamefont
  {Young}}, \bibinfo {author} {\bibfnamefont {M.}~\bibnamefont {Sarovar}}, \
  and\ \bibinfo {author} {\bibfnamefont {R.}~\bibnamefont {Blume-Kohout}},\
  }\href {http://link.aps.org/doi/10.1103/PhysRevX.3.041013} {\bibfield
  {journal} {\bibinfo  {journal} {Phys. Rev. X}\ }\textbf {\bibinfo {volume}
  {3}},\ \bibinfo {pages} {041013} (\bibinfo {year} {2013})}\BibitemShut
  {NoStop}%
\bibitem [{\citenamefont {Davies}(1974)}]{Davies:74}%
  \BibitemOpen
  \bibfield  {author} {\bibinfo {author} {\bibfnamefont {E.~B.}\ \bibnamefont
  {Davies}},\ }\href {http://dx.doi.org/10.1007/BF01608389} {\bibfield
  {journal} {\bibinfo  {journal} {Communications in Mathematical Physics}\
  }\textbf {\bibinfo {volume} {39}},\ \bibinfo {pages} {91} (\bibinfo {year}
  {1974})}\BibitemShut {NoStop}%
\bibitem [{\citenamefont {Lindblad}(1976)}]{Lindblad:76}%
  \BibitemOpen
  \bibfield  {author} {\bibinfo {author} {\bibfnamefont {G.}~\bibnamefont
  {Lindblad}},\ }\href {\doibase 10.1007/BF01608499} {\bibfield  {journal}
  {\bibinfo  {journal} {Comm. Math. Phys.}\ }\textbf {\bibinfo {volume} {48}},\
  \bibinfo {pages} {119} (\bibinfo {year} {1976})}\BibitemShut {NoStop}%
\bibitem [{\citenamefont {Knill}\ and\ \citenamefont
  {Laflamme}(1997)}]{Knill:1997kx}%
  \BibitemOpen
  \bibfield  {author} {\bibinfo {author} {\bibfnamefont {E.}~\bibnamefont
  {Knill}}\ and\ \bibinfo {author} {\bibfnamefont {R.}~\bibnamefont
  {Laflamme}},\ }\href {http://link.aps.org/doi/10.1103/PhysRevA.55.900}
  {\bibfield  {journal} {\bibinfo  {journal} {Phys. Rev. A}\ }\textbf {\bibinfo
  {volume} {55}},\ \bibinfo {pages} {900} (\bibinfo {year} {1997})}\BibitemShut
  {NoStop}%
\bibitem [{\citenamefont {Haag}\ \emph {et~al.}(1967)\citenamefont {Haag},
  \citenamefont {Hugenholtz},\ and\ \citenamefont {Winnink}}]{KMS}%
  \BibitemOpen
  \bibfield  {author} {\bibinfo {author} {\bibfnamefont {R.}~\bibnamefont
  {Haag}}, \bibinfo {author} {\bibfnamefont {N.~M.}\ \bibnamefont
  {Hugenholtz}}, \ and\ \bibinfo {author} {\bibfnamefont {M.}~\bibnamefont
  {Winnink}},\ }\href {\doibase 10.1007/BF01646342} {\bibfield  {journal}
  {\bibinfo  {journal} {Comm. Math. Phys.}\ }\textbf {\bibinfo {volume} {5}},\
  \bibinfo {pages} {215} (\bibinfo {year} {1967})}\BibitemShut {NoStop}%
\bibitem [{\citenamefont {Bravyi}\ and\ \citenamefont
  {Vyalyi}(2005)}]{Bravyi:2003tx}%
  \BibitemOpen
  \bibfield  {author} {\bibinfo {author} {\bibfnamefont {S.}~\bibnamefont
  {Bravyi}}\ and\ \bibinfo {author} {\bibfnamefont {M.}~\bibnamefont
  {Vyalyi}},\ }\href {http://arXiv.org/abs/quant-ph/0308021} {\bibfield
  {journal} {\bibinfo  {journal} {Quantum Inf. and Comp.}\ }\textbf {\bibinfo
  {volume} {5}},\ \bibinfo {pages} {187} (\bibinfo {year} {2005})}\BibitemShut
  {NoStop}%
\bibitem [{\citenamefont {Milonni}\ \emph {et~al.}(1983)\citenamefont
  {Milonni}, \citenamefont {Ackerhalt},\ and\ \citenamefont
  {Galbraith}}]{PhysRevLett.51.1108.3}%
  \BibitemOpen
  \bibfield  {author} {\bibinfo {author} {\bibfnamefont {P.~W.}\ \bibnamefont
  {Milonni}}, \bibinfo {author} {\bibfnamefont {J.~R.}\ \bibnamefont
  {Ackerhalt}}, \ and\ \bibinfo {author} {\bibfnamefont {H.~W.}\ \bibnamefont
  {Galbraith}},\ }\href
  {http://journals.aps.org/prl/abstract/10.1103/PhysRevLett.50.966} {\bibfield
  {journal} {\bibinfo  {journal} {Phys. Rev. Lett.}\ }\textbf {\bibinfo
  {volume} {50}},\ \bibinfo {pages} {966} (\bibinfo {year} {1983})},\ \bibinfo
  {note} {{Erratum Phys. Rev. Lett. \textbf{51}, 1108 (1983).}}\BibitemShut
  {Stop}%
\bibitem [{\citenamefont {Crisp}(1991)}]{PhysRevA.43.2430}%
  \BibitemOpen
  \bibfield  {author} {\bibinfo {author} {\bibfnamefont {M.~D.}\ \bibnamefont
  {Crisp}},\ }\href {\doibase 10.1103/PhysRevA.43.2430} {\bibfield  {journal}
  {\bibinfo  {journal} {Phys. Rev. A}\ }\textbf {\bibinfo {volume} {43}},\
  \bibinfo {pages} {2430} (\bibinfo {year} {1991})}\BibitemShut {NoStop}%
\bibitem [{\citenamefont {Ford}\ and\ \citenamefont
  {O'Connell}(1997)}]{Ford1997377}%
  \BibitemOpen
  \bibfield  {author} {\bibinfo {author} {\bibfnamefont {G.}~\bibnamefont
  {Ford}}\ and\ \bibinfo {author} {\bibfnamefont {R.}~\bibnamefont
  {O'Connell}},\ }\href {\doibase 10.1016/S0378-4371(97)00265-3} {\bibfield
  {journal} {\bibinfo  {journal} {Physica A}\ }\textbf {\bibinfo {volume}
  {243}},\ \bibinfo {pages} {377} (\bibinfo {year} {1997})}\BibitemShut
  {NoStop}%
\bibitem [{\citenamefont {Schaller}\ and\ \citenamefont
  {Brandes}(2008)}]{Schaller:2008aa}%
  \BibitemOpen
  \bibfield  {author} {\bibinfo {author} {\bibfnamefont {G.}~\bibnamefont
  {Schaller}}\ and\ \bibinfo {author} {\bibfnamefont {T.}~\bibnamefont
  {Brandes}},\ }\href {http://link.aps.org/doi/10.1103/PhysRevA.78.022106}
  {\bibfield  {journal} {\bibinfo  {journal} {Physical Review A}\ }\textbf
  {\bibinfo {volume} {78}},\ \bibinfo {pages} {022106} (\bibinfo {year}
  {2008})}\BibitemShut {NoStop}%
\bibitem [{\citenamefont {Fleming}\ \emph {et~al.}(2010)\citenamefont
  {Fleming}, \citenamefont {Cummings}, \citenamefont {Anastopoulos},\ and\
  \citenamefont {Hu}}]{Fleming2010}%
  \BibitemOpen
  \bibfield  {author} {\bibinfo {author} {\bibfnamefont {C.}~\bibnamefont
  {Fleming}}, \bibinfo {author} {\bibfnamefont {N.~I.}\ \bibnamefont
  {Cummings}}, \bibinfo {author} {\bibfnamefont {C.}~\bibnamefont
  {Anastopoulos}}, \ and\ \bibinfo {author} {\bibfnamefont {B.~L.}\
  \bibnamefont {Hu}},\ }\href {\doibase 10.1088/1751-8113/43/40/405304}
  {\bibfield  {journal} {\bibinfo  {journal} {J. Phys. A}\ }\textbf {\bibinfo
  {volume} {43}},\ \bibinfo {pages} {405304} (\bibinfo {year}
  {2010})}\BibitemShut {NoStop}%
\bibitem [{\citenamefont {Pekola}\ \emph {et~al.}(2010)\citenamefont {Pekola},
  \citenamefont {Brosco}, \citenamefont {M{\"o}tt{\"o}nen}, \citenamefont
  {Solinas},\ and\ \citenamefont {Shnirman}}]{Pekola:2010oj}%
  \BibitemOpen
  \bibfield  {author} {\bibinfo {author} {\bibfnamefont {J.~P.}\ \bibnamefont
  {Pekola}}, \bibinfo {author} {\bibfnamefont {V.}~\bibnamefont {Brosco}},
  \bibinfo {author} {\bibfnamefont {M.}~\bibnamefont {M{\"o}tt{\"o}nen}},
  \bibinfo {author} {\bibfnamefont {P.}~\bibnamefont {Solinas}}, \ and\
  \bibinfo {author} {\bibfnamefont {A.}~\bibnamefont {Shnirman}},\ }\href
  {http://link.aps.org/doi/10.1103/PhysRevLett.105.030401} {\bibfield
  {journal} {\bibinfo  {journal} {Phys. Rev. Lett.}\ }\textbf {\bibinfo
  {volume} {105}},\ \bibinfo {pages} {030401} (\bibinfo {year}
  {2010})}\BibitemShut {NoStop}%
\bibitem [{\citenamefont {Larson}(2012)}]{PhysRevLett.108.033601}%
  \BibitemOpen
  \bibfield  {author} {\bibinfo {author} {\bibfnamefont {J.}~\bibnamefont
  {Larson}},\ }\href {\doibase 10.1103/PhysRevLett.108.033601} {\bibfield
  {journal} {\bibinfo  {journal} {Phys. Rev. Lett.}\ }\textbf {\bibinfo
  {volume} {108}},\ \bibinfo {pages} {033601} (\bibinfo {year}
  {2012})}\BibitemShut {NoStop}%
\bibitem [{\citenamefont {Majenz}\ \emph {et~al.}(2013)\citenamefont {Majenz},
  \citenamefont {Albash}, \citenamefont {Breuer},\ and\ \citenamefont
  {Lidar}}]{Majenz:2013qw}%
  \BibitemOpen
  \bibfield  {author} {\bibinfo {author} {\bibfnamefont {C.}~\bibnamefont
  {Majenz}}, \bibinfo {author} {\bibfnamefont {T.}~\bibnamefont {Albash}},
  \bibinfo {author} {\bibfnamefont {H.-P.}\ \bibnamefont {Breuer}}, \ and\
  \bibinfo {author} {\bibfnamefont {D.~A.}\ \bibnamefont {Lidar}},\ }\href
  {http://link.aps.org/doi/10.1103/PhysRevA.88.012103} {\bibfield  {journal}
  {\bibinfo  {journal} {Phys. Rev. A}\ }\textbf {\bibinfo {volume} {88}},\
  \bibinfo {pages} {012103} (\bibinfo {year} {2013})}\BibitemShut {NoStop}%
\end{thebibliography}%

\end{document}